# An outlook of the user support model to educate the users community at the CMS Experiment

S. Malik

*University of Nebraska-Lincoln, Lincoln, NE 68588, USA*

K. Lassila-Perini

*Helsinki Institute of Physics, PO Box 64, FI-00014 University of Helsinki, Finland*

The CMS (Compact Muon Solenoid) experiment is one of the two large general-purpose particle physics detectors built at the LHC (Large Hadron Collider) at CERN in Geneva, Switzerland. In order to meet the challenges of designing and building a detector of the technical complexity of CMS, a globally distributed collaboration has been assembled with different backgrounds, expertise, and experience. An international collaboration of nearly 3500 people from nearly 200 institutes in 40 different countries built and now operates this complex detector. The diverse collaboration combined with a highly distributed computing environment and Petabytes/year of data being collected makes CMS unlike any other High Energy Physics collaborations before. This presents new challenges to educate and bring users, coming from different cultural, linguistics and social backgrounds, up to speed to contribute to the physics analysis. CMS has been able to deal with this new paradigm by deploying a user support structure model that uses collaborative tools to educate and reach out its users via a robust software and computing documentation, a series of hands on tutorials per year facilitating the usage of common physics tools, annual hands-on-learning workshops on physics analysis and user feedback to maintain and improve the CMS specific knowledge base. This talk will describe this model that has proved to be successful compared to its predecessors in other HEP experiments where structured user support was missing and the word of mouth or sitting with experts one-on-one was the only way to learn tools to do physics analysis. To carry out the user support mission worldwide, an LHC Physics Centre (LPC) was created few years back at Fermilab as a hub for US physicists. The LPC serves as a "brick and mortar" location for physics excellence for the CMS physicists where graduate and postgraduate scientists can find experts in all aspects of data analysis and learn via tutorials, workshops, conferences and gatherings. Following the huge success of LPC, a centre at CERN itself called LHC Physics Centre at CERN (LPCC) and Terascale Analysis Centre at DESY have been created with similar goals. The CMS user support model would also facilitate in making the non-CMS scientific community learn about CMS physics. A good example of this is the effort by HEP experiments, including CMS, to focus on data preservation efforts. In order to facilitate its use by the future scientific community, who may want to re-visit our data, and re-analyze it, CMS is evaluating the resources required. A detailed, good quality and well-maintained documentation by the user support group about the CMS computing and software may go a long way to help in this endeavour.

## 1. Introduction

The CMS detector [1] designed by an international collaboration of 3000 members is currently exploring the energy region where physicists believe they will find answers to the central questions at the heart of particle physicists. These key questions include what gives mass to the elementary particles, why is the universe predominantly matter and not antimatter, are there deeper symmetries beyond the Standard model etc. The journey to these answers has gained momentum with CMS publishing its 100th paper recently in less then two years of data taking. The superb performance of the LHC, specifically, the CMS detector, together with an enormous and well-coordinated analysis effort by engaging the collaboration on many frontiers, are primarily responsible for this milestone. Besides many key factors behind this success lies a robust computing infrastructure in place accompanied by a User Support model. With over two hundred institutes involved from forty countries around the world, the distributed nature of the collaboration makes things complicated. The users need to understand the complex CMS computing environment and the physics tools as a pre-requisite task in order to do the physics analysis. The CMS User Support engages the entire collaboration in this important task. This global task requires overcoming large distances and large time differences. This paper describes the goals, challenges, working strategy and activities of the User Support [2], the growth of LHC Physics Centers as regional hubs for the users to have face-to-face experience and possible role of User Support in preservation of CMS data.

## 2. Challenge

The CMS user community comes from different social, cultural and linguistic backgrounds that spread over large distances and large time differences. The distributed nature of CMS collaboration and computing resources presents a new paradigm for the users in order to perform physics analysis. The large size of the collaboration and the long lifetime of high-energy physics (HEP) experiments has brought in a need for an organized effort to support and engage users in meeting their physics needs. The model where the collaboration size is relatively smaller than CMS and people could do physics analysis efficiently only at the hub of detector activity is not very practical. The financial and logistic constraints may not permit everyone to travel frequently or for extended periods to CERN. In addition, the complex computing environment and non-trivial tools needed to do physics add to learning curve of a user to become fully equipped to do physics analysis. This scenario poses a new, unique and a big challenge to support the physics needs of this community.

The full physics potential of the CMS detector can be realized only if users quickly come up to speed to contribute towards physics analysis. This is possible only by learning about the CMS computing environment and the available physics and software analysis tools. The CMS Experiment has put in a significant effort to organize a User Support structure that plays a vital role to help the collaborators achieve the above goal. To engage a big collaboration like CMS in physics analyses activity, the User Support uses the available collaborative tools [3] supported by CERN. It provides a general physics support by maintaining, supervising and improving the CMS software and computing user documentation in the form of hundreds of web wikis, conducts dozens of tutorials year around on the tools needed as a pre-requisite to perform physics analyses and answers user questions via emails or in person.

## 3. Organization

The CMS Experiment is organized into several tasks for efficient management in several areas. Among them is the CMS Computing [4] that is further divided into the Data, Analysis and Facility Operations, and the User Support. They are geared towards helping users analyze data and produce physics results. The Data Operations team gets the data out of the CMS detector (online system), processes and re-processes and distributes them for analysis all over the world to different computing centers. Analysis Operations focus on the operational aspects of enabling physics data analysis at the globally distributed computing centers worldwide by managing data movement, access and validation. The Facility Operations provide and maintain a working distributed computing fabric with a consistent working environment for the Data Operations and the users. The CMS User Support provides the general computing software and physics support structure to enable users to do physics analysis.

The CMS User Support is lead by two co-conveners. There is a very small number of dedicated personal doing User Support and therefore the manpower is drawn from the CMS user community itself. The users who are expert in different physics analysis tools help new users learn and master them. They in turn become experts and provide feedback to further improve the instruction material and style. These new experts get recruited to help the User Support effort further. In this way the cycle of teaching and learning goes on in a self-sustained manner. This mode of operation is a clear manifestation of the big collaborative spirit without which it is not possible to run such a diverse and globally distributed community as CMS.

## 4. Collaborative Tools

The goal of the User Support is to engage the collaboration in learning the tools needed to perform the physics analysis and bring them up to the speed to contribute to the same. It is highly desirable to distribute the expertise besides CERN to other institutions. To meet these challenges the User Support uses the collaborative tools available and supported by CERN. And does not take part in developing any collaborative tools. However, the User Support activities do serve as a very active test bed for these tools. It maintains and supervises the CMS documentation and periodically organizes tutorials on learning computing and physics tools and workshops on physics analyses. It answers general computing and physics questions via emails and online chat sessions. The specific questions go to the respective discussion forums and get answered in a collaborative manner by the experts. Whereas tools like Indico [5], EVO (Enabling Virtual Organizations) [6], Twikis [7] and Hypernews [8] enable it to communicate with the users, for a dedicated purpose, like setting up a tutorial forum, it uses espace Sharepoint technology [9]. The usage of these tools is described below.

Indico is web-based tool used to schedule and organize events, conferences, meetings, workshops with sessions and contributions. The tool also includes an advanced user delegation mechanism; allow paper reviewing, archival of conference information and electronic proceedings. The User Support uses it to display the agenda for a tutorial, upload presentations and video recordings of the tutorials. EVO, the virtual conferencing tool, is heavily used by CMS and other LHC experiments. The User Support utilizes EVO to enable successful participation of the remote collaborators during the tutorials. The video/audio recording feature of EVO is used to record the tutorials and the recordings are uploaded to the Indico agenda.

CMS uses web twikis for the purpose of its knowledge base and documentation management system. The flexibility to edit and modify a twiki by any CMS user and instantly publish it using only a web browser (no programming required) makes them very suitable for a collaborative environment. The User Support also uses, espace, an online collaboration tool based on Microsoft SharePoint technology, to prepare and conduct tutorials. This is used to plan and build a course workflow that goes from announcing a tutorial to getting feedback. It serves as central platform that has links to the tutorial announcement, Indico agenda for the tutorial lectures and exercises, forms to submit exercises, discussion forum and the feedback form.

Hypernews (HN) is a discussion management system used by all LHC experiments including CMS. It bridges the use of e-mail and forums. The user may subscribe to forums to be informed by e-mail or look at the forums of his choice via the web interface. He may reply by e-mail or through the web interface. Discussions are grouped into Forums and threads, and Forums are in turn grouped into Categories. Currently, CMS has about 20 such Hypernews categories with about 700 forums. HyperNews is used to send announcement about tutorials and documentation review and to respond to questions/comments from the users about documentation.

## 5. Activities

In this section we will describe how do we use the collaborative tools CMS community-wide to spread the knowledge of analysis tools and related communication in order to successful do physics analysis. In particular these tools are used for documentation, tutorials, software reference and presentations.

### 5.1. Documentation

CMS has about few hundred twikis that need to be maintained by the User Support as part of CMS documentation. The usability of this documentation has been tested [10]. The challenge of such an enormous amount of information is to track the information that may be stale or outdated. To deal with this and improve the performance of the twiki usage, a reminder service of twiki pages has been tested and deployed. As a result, each twiki responsible gets a reminder about the pages that he or she maintains and is encouraged to remove the outdated information. To reduce effort on part of the responsible, the reminder service, in addition, offers a possibility of a delete request. The twiki user only needs to select the topics to be deleted and they will be deleted centrally.

The User Support manages and periodically reviews the structured documentation described below. Many topics are also reviewed by the end-users to make the information as best as possible.

WorkBook [11] serves as the first entry point for a user on how to go about approaching physics analysis. It in the form of about a dozen chapters spread over about 100 twiki pages. The twiki page on each topic has a responsible person associated with it. The WorkBook covers all aspects to help and equip a user jump-start physics analysis. One learns about how to get computer accounts and learn CMS computing framework, common physics tools and examples on advance analyses.

The Software Guide (SWGuide) [12] has details of each domain belonging to the CMS software. It contains details about the data formats, software framework and physics analysis software. There are about 1350 twiki pages and each topic has a responsible. For example, if one looks at the WorkBook topic on how to use Physics Analysis Toolkit (PAT) [13], one can find more details on its concept and coding structure in the SWGuide.

CMS uses Doxygen [14] as a tool to generate and maintain reference manual for data formats and CMS software. With this tool one can generate the documentation from the software code for each release of the CMSSW. The User Support has provided software scripts to better integrate the reference manual with the rest of the documentation. The user, while reading the document, can easily refer to the actual code. The software manual contains cross-links to the main documentation in the SWGuide and lists all the packages connected to each software domain.

### 5.2. Tutorials

The tutorials [15] are organized throughout the year on common physics analysis tools. They are given by the experts and rely on the collaborative spirit. The number of attendees in a given tutorial is limited by the practical factors such as space in the conference room and number of tutors available. Some topics on which the tutorials are held are described below along with their importance towards accomplishing physics analysis. The tutorials are mostly held at CERN but also frequently at other collaborative labs like FNAL (Fermi National Accelerator Lab), DESY (Deutsches Elektronen-Synchrotron) etc. The presentations and video recordings of the tutorials are uploaded to Indico as a persistent storage for the users to view later. Frequent tutorials ensure an up to date documentation. These tutorials are described in the rest of the sub-section.

The CMS software framework is very complex. It is essential to work on any computing aspect of the CMS Detector Studies or Physics Analysis. It uses configuration files scripted in Python language to configure the initial settings for computer programs. Thus having CMSSW tutorial along with the usage of Python is required. This tutorial is now a pre-requisite for part Physics Analysis Toolkit (PAT) tutorial described below.

Analyzing physics data at LHC experiments is a complicated task involving multiple steps, sharing of expertise, cross checks and comparing different analyses. To maximize physics productivity, the CMS experiment at LHC has developed a special collection of analysis tools called the PAT (Physics Analysis Toolkit). A comprehensive training program was designed and setup on using PAT software as an integral key part of the analysis of data from the CMS experiment. Learning CMS software framework and python as the configuration language are also a part PAT tutorial as pre-requisite exercises. A typical number of participants attending PAT tutorials is around 25, though some additional participants attend the tutorials remotely.

To keep CMS community abreast with statistical tools and modeling of the physics event distribution in data analysis, a combined RooStats [16] and RooFit [16] tutorial is organized. One learns to implement techniques like - frequentists, Bayesian and likelihood based methods for statistical calculations.

Analysis of CMS data is performed in a distributed way using grid infrastructure. CMS Analysis Remote Builder (CRAB) [17] is a utility to create and submit CMSSW jobs to the distributed computing resources. It interacts with the local user environment, the CMS Data Management services and with the Grid middleware. It is essential to learn this tool via a tutorial in order to successfully perform one's data analysis.

The 3D-accurate display to visualize proton-proton collisions is very essential to understand the physics and the outcome of the physics event. The CMS event display, called Fireworks [18], is specialized for this purpose. The collision data is presented via user-friendly graphical and textual views. This tool is easy to install and use. To benefit maximum to one's need does require some detailed coaching in the form of tutorial.

To give orientation about CMS, series of lectures are organized every summer for the CERN summer students and the new comers to the CMS. These lectures give them an overview of CMS detector, physics and software.

Holding physics analysis workshops has also been a great way of introducing new members of the collaboration - students, post docs and faculty - to CMS software and data analysis. User Support organizes workshop, called CMS data Analysis School [19], on analysis problems on collision data as well as data simulating the conditions at LHC. The experienced volunteers from the CMS facilitate team of tutors and lecturers. Teams of participants and attendees on the order of about 100 present their simulated physics measurements in "conference" at the end of the workshop. The material prepared for this workshop serves as a persistent resource for CMS after the close of the Workshop.

User Support also provides in-person help where a user can walk in and ask questions during specific office hours.

## 6. User Feedback

The feedback from users and their help is the key to the success of the user support. For every PAT tutorial and analysis school that is conducted, a feedback form is supplied to the participants for suggestions and improvements. This process has helped the tutorials to grow and mature in terms of instructional material, structure and organization.

Last year, the CMS computing project conducted a survey to see how CMS worked for the users with respect to their computing and analysis needs. This survey included specific questions about the User Support. About 120 participants took part in it. The survey indicated that the users find SWGuide and WorkBook as very valuable as shown in Figures 1 and 2. The users also felt that more topics for tutorials should be added. The users prefer twiki pages for self-study and HN to direct their questions. Getting help from these resources, around 40% users said that it took them 1-2 weeks to get started with their analysis. This is a very big success given the complexity of physics analysis, tools and software.

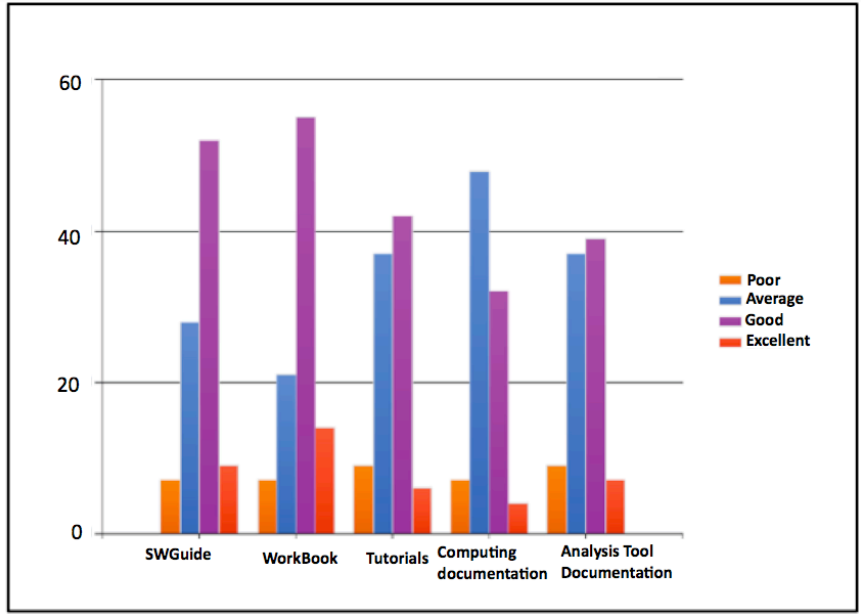

Figure 1: User rating for the documentation

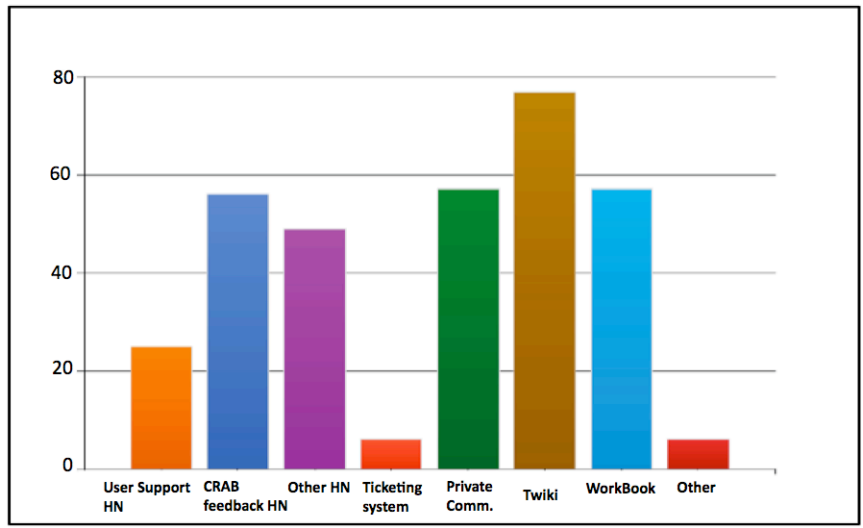

Figure 2: Support channel used by the users

## 7. Growth of LHC Physics Centers

The scale and complexity of CMS experiment are unprecedented. The success of the early LHC physics program, achieved so quickly after startup, is the product of a globally distributed scientific effort of an unprecedented scale, suggesting a new paradigm in scientific collaboration; one in which significant numbers of scientists are no longer co-located at the host lab, but instead are at individual institutions or grouped regionally. Many CMS colleagues are engaged but many are not. This situation may be particularly important for CMS, as we are a big collaboration. To be engaged in a global collaboration requires overcoming large distances and large time differences. Modern collaborative tools are powerful enablers - but for most face-to-face still remain indispensible. In this scenario, the role and efficacy of a remote regional center can hardly be overemphasized. This new paradigm in scientific collaboration has not only lead to the organized and central User Support but also growth of LHC Physics Centers worldwide.

For a typical physics group at a collaborating university, there are several paths to be plugged into contributing towards physics analysis. They can place a team of physicists and students at CERN or at home institute or at an LHC Physics Centre (regional centre). Most groups choose a combination of the above and in this way a significant numbers of scientists, no longer co-located at the host lab can contribute being at individual institutions or grouped regionally. Today three such centres of excellence exist worldwide – the LPC at Fermilab [20], U.S.A, the LPC Centre, called LPCC [21] at CERN (host lab), Switzerland and Terascale [22] at DESY, Germany. The LPC at Fermilab came into existence over five years ago. Following its unprecedented success, the centres at CERN and DESY came into existence. The LHC Physics Centers serve as venues for tutorials, talks by HEP experts, help from experts, analysis workshops etc. The role and efficacy of a remote regional center is vital for CMS.

The LHC Physics Center (LPC) at Fermilab is a regional centre of the CMS collaboration in U.S.A. The LPC serves as a resource and physics analysis hub primarily for the seven hundred US physicists in the CMS collaboration. It offers a vibrant community of CMS scientists from the US and overseas who play leading roles in analysis of data, definition and refinement of physics objects, detector maintainance, and in the design and development of the future detector upgrade. There is close and frequent collaboration with the Fermilab theory community. The LPC provides outstanding computing resources and software support personnel. The proximity of the Tier-1 and the Remote Operations Centre allow critical and real time connections to the experiment. The LPC offers educational workshops in data analysis, and organizes conferences and seminar series. The LPC, though regional centre, is not restricted to CMS users from U.S.A. Any CMS member worldwide is welcome to come and work here. There are a lot of students from non-US CMS institutes who work at LPC, for example India, Turkey, China and South America.

The LPCC at CERN epitomizes a set of initiatives in support of the LHC physics program. These range from the organization of Workshops, to the support of the development of event generators and other physics tools, to the development of corresponding web pages. It is intended as a portal into LHC physics resources for the whole HEP community. These initiatives include (a) the LHC working groups that provides a common forum for discussion among the LHC experiments and theoretical physicists, (b) center for events like workshops, LHC Physics Day, EP/PP/LPCC seminar, LPCC students' lecture and LHC reports, (c) physics analysis tools and (d) theory contacts: the list of theorists at CERN that one can contact for physics questions, simulation support, BSM (Beyond Standard Model) ideas to be pursued etc.

In order to optimally place German particle physics in a global research environment, the new structure of the Helmholtz Alliance 'Physics at the Terascale' was founded, bundling the German activities in the field of high-energy collider physics. It is a network comprising of all German research institutes working on LHC experiments, a future linear collider or the related phenomenology - 18 universities, two Helmholtz Centres and one Max Planck Institute. The Alliance works on topics like the development of new accelerator and detector technologies, methods of data analysis, development of theoretical models and methods and development of the relevant computing infrastructure. The Alliance acts as a collective tool for a more effective collaboration, in particular between experimentalists and theorists. A common infrastructure has been created where all partners of the Alliance contribute to and use these infrastructures for specific research projects, for example, National Analysis Facility (NAF) that provides training, support and own contributions for all analysis-related issues, currently focusing on Monte Carlo generators, parton distribution functions, statistics tools and the interpretation of physics beyond the Standard Model of particle physics. The other example is the Virtual Theory Institute that (VTI) holds regular seminars starting in June 2008. These are broadcast via EVO and all Alliance members can participate.

In the following section we would describe in detail about the success model of LPC at Fermilab.

### 7.1. LHC Physics Center (LPC) at Fermilab

#### 7.1.1. The LPC structure

A board called LPC Management Board (LPCMB) manages the organization of LPC. The board charts its evolution for the benefit of CMS and focuses on strategy and tactics. It comprises of CMS Physics Coordinators, representatives from US Universities on CMS and Fermilab, USCMS Research Program and LPC coordinators. The LPCMB by virtue of its composition ensures that the LPC is well coordinated with the rest of the CMS. Specifically, the LPCMB directs the LPC program of work, draws up policies and provides the forum for close coordination with activities in CMS. Through this process the LPC maximally benefits CMS. The LPCMB is best known for physics organization structure of the LPC, the LPC Fellows program and selection of the fellows. The day-to-day operations of LPC are taken care of by the two LPC coordinators.

#### 7.1.2. Engaging the CMS Community

CMS physics and physics object activities with LPC involvement span the complete range of the CMS program where many CMS conveners are based at LPC. LPC supports young CMS leaders through its LPC fellows program and LPC Guest & Visitor Program - short and long stays. These are not restricted to US members. It support students, postdocs & faculties and is designed to help build and maintain the strength of CMS. The LPC fellow program was started in January of 2011 and supports 14 young and midcareer emerging leaders on CMS. The focus of physics of these fellows is Higgs, SUSY and Exotica. Fellows use resources of LPC (physical & human) to make a major contribution to CMS physics or physics object refinement. It serves as a unique opportunity for the young scientists to build independent research program.

### 7.1.3. Engaging the HEP Experts

The LPC has been at forefront of CMS activities by organising week long workshops on physics tools and data analysis and engaging experts in the area by inviting them for a week to give several lectures/seminars, hold "office hours", chat with locals in person or by EVO with CMS. LPC has organized over half a dozen workshops in the last 4 years to equip the physicists with the tools (analysis toolkit, grid usage, event displays etc.) needed to do physics analysis. It also organised CMS Data Analysis School (CMSDAS) in January 2011, the very first of its kind. The school was designed to help CMS physicists from across the collaboration to learn, or to learn more, about CMS analysis and thereby to participate in significant ways to the discovery and elucidation of the new physics. The innovative classes allowed the students, in some cases with zero experience, to have hands on experience with real data physics measurements that CMS published just days before and then make them more precise by searching for new processes that the collaboration hasn't done yet. The students were expected to continue the work on the measurements they started and then see it through to publication after the school was over. It provided opportunity to the new members to meet in person with a lot of the experts and was a very good way for bringing a new generation of people into the experiment. 90% of the time during the school was devoted to the hands on analysis work with only 10% physics talks. Over 60 students and 60 facilitators took part in the school. Of these 20% participants came from outside the US as far as from Brazil, Korea and Europe. The school was a huge success and got the CMS management further interested and there is now a plan in the works to hold this school at an annual frequency across the CMS institutions in different continents. An interest to hold similar schools interest has already been expressed by DESY, INFN Pisa, Asia (Taiwan) and US-ATLAS. The material presented for the school is part of the CMS WorkBook and is supported and maintained by the CMS Analysis Tools and User Support team.

### 7.1.4. Engaging Theory and Experiments

LPC is leading a concerted effort to partner with sister LHC centers along with engaging theory (CTEQ collaboration) and other HEP experiments – ATLAS and Tevatron. Recently, two workshops on Standard Model benchmarks at the high-energy colliders (Tevatron, LHC) have been organized at Fermilab and DESY. The next one is planned at an ATLAS analysis center.

### 7.1.5. LPC contribution to physics publications

CMS physics and object activities with LPC involvement span the complete range of the CMS program – SUSY, Higgs, Exotica, b-quark tagging to name a few. Many of the CMS conveners are based at the LPC. A measure of the LPC contribution to the CMS physics and its success can be gathered from the fact that, of the over 100 publications so far, by the CMS experiment, LPC has a direct contribution in a third of these.

## 8. Data Preservation at CMS

The need and value in the data preservation [23] of HEP experiments has been brought forth in the last couple of years. Typically, an HEP experiment lasts for a few years during which their unique data are collected, analyzed and published. Once the experiment stops collecting the data, the collaboration and the knowledge base for that experiment disintegrates fast, as people move on to the other experiments and careers. However, developments of the new theories and the new techniques from the simulations and experiments is a continuous intellectual process and may require verification and revisit of the HEP data collected by an HEP experiment. This thought is also gaining importance as the energy and luminosity frontiers expand in the future (CMS/ATLAS at LHC) and new ideas and discoveries are brought forth. But a re-look and re-analysis of the data presents significant challenges. For example, dataset collected by the experiments that have come to a close (e.g. BABAR at SLAC) or would soon close (CDF/D0 at the Fermilab Tevatron) are unique for their collisions energy and those conditions might never be repeated in future given the high costs involved. The corresponding software and analysis techniques become non-usable as experts move on and new technology and computing techniques supersede. The computing power might also not be available as the commitment from the host lab has lived its duration. Given the above scenario, time has come to confront the question on how to preserve and access this unique, rich and valuable data in future. To address this specific issue in an organized and systematic way, the Study Group on Data Preservation and Long Term Analysis in High Energy Physics was formed at the end of 2008 with the aim to clarify the objectives and the means of preserving data in high-energy physics

## 8.1. CMS Data preservation policy and implementation

The CMS collaboration is committed to preserve its data and allow its use by a wide community: collaboration members long after the data are taken, experimental and theoretical HEP scientists who were not members of the collaboration, educational and outreach initiatives, and citizen scientists among the general public. The lower-energy and lower-luminosity LHC data taken already at 0.9, 2.36 and 7TeV will never be repeated, and their preservation and preparation for later re-use, has to be addressed urgently. Meeting this challenge would also offer a unique way to stress test and evaluate the entire preservation, re-use and (open) access concepts for the CMS data. In this spirit, the CMS is currently beginning to engage itself in this task and is working on a general policy document on data preservation and a detailed implementation plan. The policy would describe the CMS principles of data preservation, its reuse and open access, as well as the identifying the responsible personal and their roles in each of these phases. It should be realised and emphasized that in order to fully exploit all the reuse of the data, an immediate commitment and continued resources would be needed in CMS.

The implementation plan would have the details on how to identify each and every step involved towards data preservation and reuse. It would follow the four levels described by the DPHEP working group in [24] as a reference for the CMS Policy. It should assess the current situation at CMS in terms of each of these levels. It should also describe the evolution of data preservation in short-term future as well as during the entire data-taking period of the experiment and beyond.

## 8.2. User Support and Data preservation

The data preservation effort encompasses CMS software, analysis code and techniques, and raw and reconstructed data. A very first step towards this endeavor would be to have a robust and up-to-date documentation about each of them. The CMS WorkBook, SWGuide and Software reference documentation, described above, has all the basic as well as expert level information about these. The CMS User Support is responsible and periodically reviews these. There is a responsible identified to maintain and update each topic. Given the numerous tutorials, analysis schools held by User Support team, we have trained over ~500 CMS users over last 2 years. This number is expected to grow cumulatively each year. The methods and technology that is used to train CMS users in house can easily be extended to include and train non-CMS community that might be interested in our science and revisiting our data.

## 9. Conclusions

The CMS User Support model has proved to be successful compared to its predecessors in other HEP experiments. It has put systematic and organized effort to equip users do physics analysis. It has made impact in awareness and usability of the structured documentation suite, facilitating the usage of common physics tools. It relies on the collaborative effort in maintaining, contributing and improving the documentation and holding tutorials. The user feedback is essential to maintain and improve the CMS specific knowledge base. The collaborative spirit is the key to its success. To engage as much collaboration as possible in doing physics analysis has given birth to LHC Physics Centres around the world. They act as catalyst for physics learning and mutual exchange of ideas leading to discoveries. The User Support activities are also very well aligned with the new and big effort of data preservation at CMS and can further guide it.